\documentclass[sigconf, nonacm]{acmart}
\usepackage{amsmath}
\usepackage{xcolor}
\usepackage{colortbl}
\usepackage{multirow}
\usepackage{subfigure}
\usepackage{todonotes}

\usepackage{algorithm}
\usepackage[noend]{algpseudocode}
\usepackage{makecell}

\algrenewcommand\algorithmicrequire{\textbf{Input:}}
\algrenewcommand\algorithmicensure{\textbf{Output:}}

\newcommand\vldbdoi{XX.XX/XXX.XX}
\newcommand\vldbpages{XXX-XXX}
\newcommand\vldbvolume{17}
\newcommand\vldbissue{1}
\newcommand\vldbyear{2024}
\newcommand\vldbauthors{\authors}
\newcommand\vldbtitle{\shorttitle} 
\newcommand\vldbavailabilityurl{URL_TO_YOUR_ARTIFACTS}
\newcommand\vldbpagestyle{plain} 
 \newcommand{\showDOI}[1]{\unskip}

\begin{document}
\title{KnobCF: Uncertainty-aware Knob Tuning}


\author{Yu Yan}
\affiliation{%
    \institution{Harbin Institute of Technology}
    \streetaddress{92 West Dazhi St}
    \city{Harbin}
    \state{Heilongjiang}
    \country{China}
}
\email{yuyan@hit.edu.cn}

\author{Junfang Huang}
\affiliation{%
    \institution{Harbin Institute of Technology}
    \streetaddress{92 West Dazhi St}
    \city{Harbin}
    \state{Heilongjiang}
    \country{China}
}
\email{twiherhol@gmail.com}

\author{Hongzhi Wang}
\affiliation{%
    \institution{Harbin Institute of Technology}
    \streetaddress{92 West Dazhi St}
    \city{Harbin}
    \state{Heilongjiang}
    \country{China}
}
\email{wangzh@hit.edu.cn}

\author{Jian Geng}
\affiliation{%
    \institution{Harbin Institute of Technology}
    \streetaddress{92 West Dazhi St}
    \city{Harbin}
    \state{Heilongjiang}
    \country{China}
}
\email{gengj@stu.hit.edu.cn}

\author{Kaixin Zhang}
\affiliation{%
    \institution{Harbin Institute of Technology}
    \streetaddress{92 West Dazhi St}
    \city{Harbin}
    \state{Heilongjiang}
    \country{China}
}
\email{21B903037@stu.hit.edu.cn}

\author{Tao Yu}
\affiliation{%
    \institution{Harbin Institute of Technology}
    \streetaddress{92 West Dazhi Street}
    \city{Harbin}
    \state{Heilongjiang}
    \country{China}
    \postcode{150001}
}
\email{21B903056@stu.hit.edu.cn}

\begin{abstract}
The knob tuning aims to optimize database performance by searching for the most effective knob configuration under a certain workload. Existing works suffer two significant problems. On the one hand, there exist multiple similar even useless evaluations of knob tuning even with the diverse searching methods because of the different sensitivities of knobs on a certain workload. On the other hand, the single evaluation of knob configurations may bring overestimation or underestimation because of the query uncertainty performance.  To solve the above problems, we propose a decoupled query uncertainty-aware knob classifier, called \textsf{KnobCF}, to enhance the knob tuning. Our method has three significant contributions: (1) We propose a novel concept of the uncertainty-aware knob configuration estimation to enhance the knob tuning process. (2) We provide an effective few-shot uncertainty knob estimator without extra time consumption in training data collection, which has a high time efficiency in practical tuning tasks. (3) Our method provides a general framework that could be easily deployed in any knob tuning task because we make no changes to the knob tuners and the database management system. Our experiments on four open-source benchmarks demonstrate that our method effectively reduces useless evaluations and improves the tuning results. Especially in TPCC, our method achieves competitive tuning results with only 60\% to 70\% time consumption compared to the full workload evaluations.

\end{abstract}

\maketitle

\pagestyle{\vldbpagestyle}
\begingroup\small\noindent\raggedright\textbf{PVLDB Reference Format:}\\
\vldbauthors. \vldbtitle. PVLDB, \vldbvolume(\vldbissue): \vldbpages, \vldbyear.\\
\href{https://doi.org/\vldbdoi}{doi:\vldbdoi}
\endgroup
\begingroup
\renewcommand\thefootnote{}\footnote{\noindent
This work is licensed under the Creative Commons BY-NC-ND 4.0 International License. Visit \url{https://creativecommons.org/licenses/by-nc-nd/4.0/} to view a copy of this license. For any use beyond those covered by this license, obtain permission by emailing \href{mailto:info@vldb.org}{info@vldb.org}. Copyright is held by the owner/author(s). Publication rights licensed to the VLDB Endowment. \\
\raggedright Proceedings of the VLDB Endowment, Vol. \vldbvolume, No. \vldbissue\ %
ISSN 2150-8097. \\
\href{https://doi.org/\vldbdoi}{doi:\vldbdoi} \\
}\addtocounter{footnote}{-1}\endgroup

\ifdefempty{\vldbavailabilityurl}{}{
\vspace{.3cm}
\begingroup\small\noindent\raggedright\textbf{PVLDB Artifact Availability:}\\
The source code, data, and/or other artifacts have been made available at \url{https://github.com/AvatarTwi/KnobCF}.
\endgroup
}
\section{Introduction}
\label{sec:intro}
The knob tuning task is typically defined as the process of optimizing database performance by searching for the most effective knob configuration~\cite{zhao2023automatic}. The knob configuration directly determines the resource partition, operator execution, etc., significantly influencing the workload performance of the database system~\cite{van2017automatic}. Existing solutions for knob tuning typically consist of two major parts: (i) \textbf{Configuration Recommendation:} searches knob configuration according to heuristic rules~\cite{zhu2017bestconfig, sullivan2004using}, Bayesian Optimization Models~\cite{kanellis2022llamatune}, Reinforcement Learning Agents~\cite{zhang2019end, li2019qtune, wang2021udo}, etc. (ii) \textbf{Workload Evaluation:} evaluates the workload performance based on selected knob configuration. Among these steps, the most time-consuming part is the workload evaluation. For example, in the open source benchmark, TPCH, we observe that CDBtune~\cite{zhang2019end} spends over 80\% of time evaluating the performance of knob configurations. Thus, the efficiency of the evaluation is crucial for the knob tuning results. 

We observe two significant problems in the workload evaluation of knob tuning that cause the efficiency issue. 

\begin{figure}[h!]
    \centering
    \includegraphics[width=0.99\linewidth]{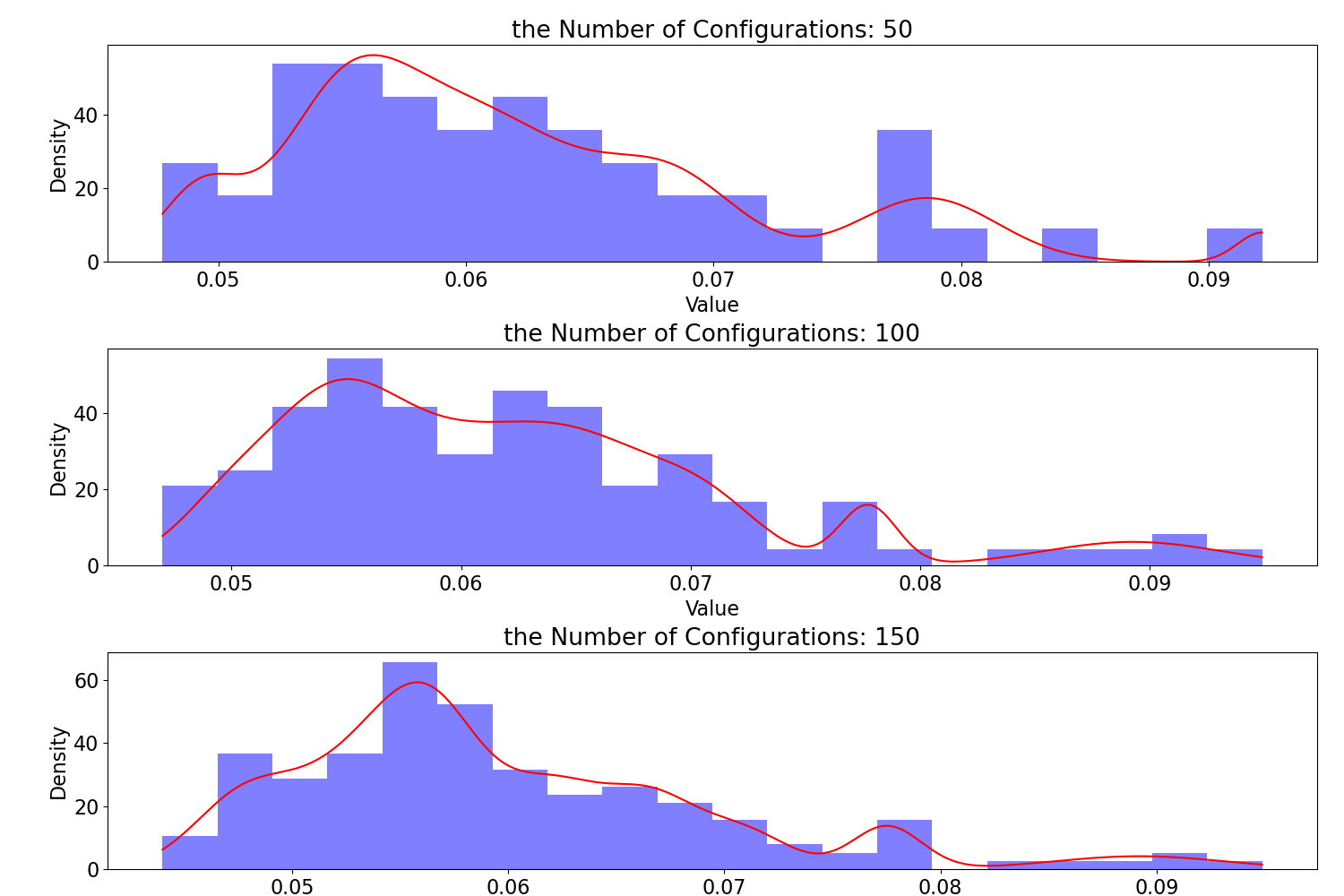}
    \caption{The probability density curve of query latency distribution after different tuning iterations on YCSB-a.}
    \label{fig:uncertain}
\end{figure}

First, although existing work makes the best efforts to recommend diverse knob configurations, multiple similar even useless evaluations still exist in the knob tuning process, leading to low time efficiency. Taking Figure~\ref{fig:uncertain} as an example, we present the probability density curve of the query latency distribution on the YCSB-a knob tuning process (we utilize DDPG~\cite{zhang2019end} as the knob tuner). We observe that the query latency distribution rarely changes with the tuning iterations, indicating that there exists a large number of iterations that obtain similar even same query latency with the former iterations. This phenomenon is caused by the different sensitivities of database knobs on a certain workload, i.e., the same modification on sensitive knobs brings more huge performance changes than the non-sensitive knobs. Thus, the tuning process contains many useless evaluations even with various configuration search strategies. 

Second, there exists a natural contradiction between time efficiency and accuracy in evaluating the workload performance of knob configurations. 
On the one hand, to save evaluation time consumption, some methods~\cite{zhao2023automatic, kanellis2022llamatune} utilize a one-time evaluation as the estimated performance of knob configurations. However, due to the uncertain workload execution, one-time evaluation may result in overestimation or underestimation, leading to wrong knob tuning decisions. On the other hand, multi-time evaluations~\cite{wu2014uncertainty} will bring more accurate and robust knob tuning while causing large time consumption. And in large database systems~\cite {zhang2021restune}, it may be better to continue to run with a worse configuration instead of executing multiple evaluations for knob tuning. 

In general, the above two problems could be concluded as how to implement an efficient evaluation for knob configurations. To address the first problem, we consider modeling the performance distribution of knob configuration. To address the second problem, we combine the uncertainty distribution with performance modeling. 
Specifically, we propose the concept of uncertainty-aware evaluation for knob tuning, i.e., we aim to model the uncertain performance distribution of knob configurations. Importantly, knob tuning is a multi-objective optimization problem, which has multiple evaluation metrics, such as workload execution time, memory usage, disk IO usage, etc. In this paper, we limit our focus on predicting the workload execution time (Throughput) of knob configurations based on our uncertainty-aware knob estimator. In addition, the uncertainty of other evaluation metrics is interesting to study in the future.  

\begin{figure*}[ht]
    \centering
    \includegraphics[width=0.80\linewidth]{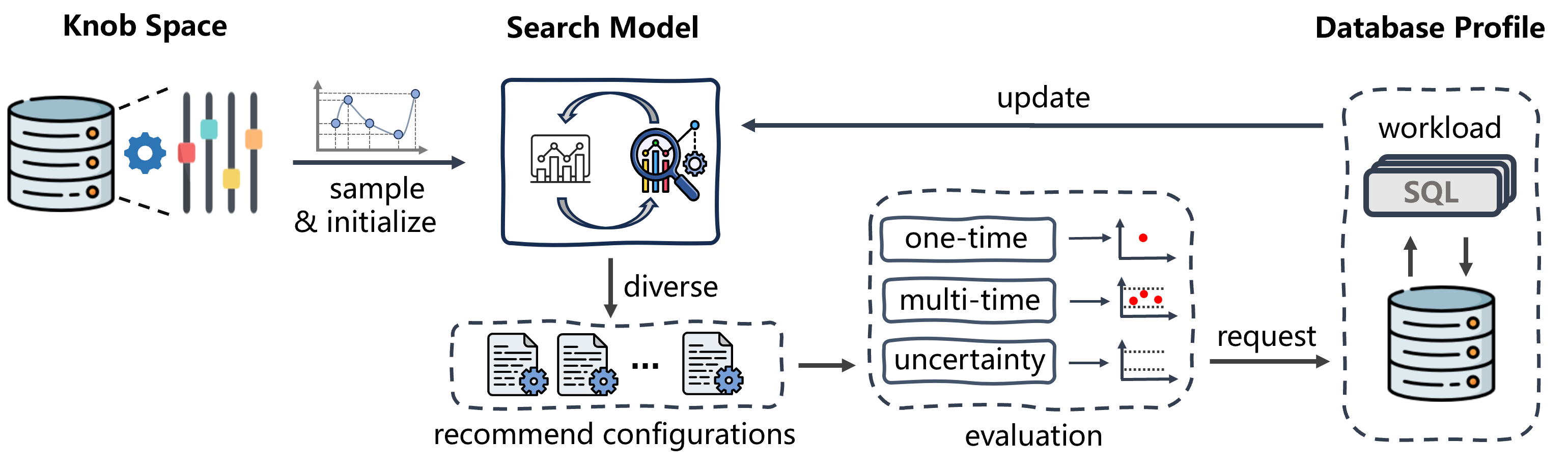}
    \caption{The Knob Tuning Process.}
    \label{fig:tuning}
\end{figure*}

For execution uncertainty, some existing works~\cite{wu2014uncertainty} have proposed some machine learning estimators to model the query uncertainty distribution. However, these methods focus on underlying uncertain cardinality sampling while lacking the consideration of knob configuration. Moreover, to our best known, existing works have no formal definition of the uncertainty distribution under knob configurations. Even though the uncertainty of knob configuration could enhance the evaluations of knob tuning, it brings several challenges:

(1)\textbf{High-Dimensional Knob Candidate Space:} Even after applying knob filtering techniques~\cite{kanellis2020too}, the space of potential knob configurations remains high-dimensional. This requires multiple query executions to gather sufficient training data on the uncertain distribution. (2)\textbf{Diverse Queries:} Each query may have a unique structure, different table schema, and various indexes making it difficult to design a universal feature representation which is the basis for designing a model transfer mechanism. (3)\textbf{Time Efficiency:} It is especially crucial for knob tuning tasks designed for changed workloads to construct a model transfer mechanism. This means we should avoid extensive pre-training for workload adjustments and design the model transfer mechanism to reduce training time consumption. 

\noindent\textbf{Our Approach.} To address these challenges, we propose a query uncertainty-aware \underline{knob} \underline{c}lassi\underline{f}ier called \textsf{KnobCF} to enhance the knob tuning. Our key observation is that a considerable portion of knob configurations have similar performance distribution with evaluated knob configurations. In these cases, it is unnecessary to reevaluate these knob configurations. Instead, we estimate the performance of new knob configurations based on historical evaluations, to largely reduce the useless evaluations.  Also, due to the uncertainty modeling, our method could alleviate the overestimation and underestimation problems caused by single-point estimation. In addition, our uncertainty-aware estimator is easy to deploy in any knob runners, which is decoupled with the knob optimizers. 

Specifically, our approach addresses the above challenges in three aspects. 

(1) \textbf{Problem Definition:} Instead of simply defining a regression task to predict the uncertainty distribution of single knob configurations, we propose an innovative problem definition, a joint distribution uncertainty classifier, to predict the joint distribution classification label of knob configurations. On the one hand, we utilize the joint distribution of knob configurations instead of the single knob distribution, largely reducing the time overhead of collecting training data. On the other hand, compared with predicting regression distribution statistics, the design of classification label prediction can efficiently prevent the model from being affected by different scales of various tuning tasks.

(2) \textbf{Feature Representation:} The feature representation is a fundamental task for effective model training and inference. To process the diverse queries, we combine the graph convolutional network~\cite{hilprecht2022zero} and knob importance~\cite{zhang2023openfe} to encode the query plan into a fixed embedding vector, which could easily gather sufficient training data to obtain high-quality embedding vectors. Also, we design the universal knob encoding method by utilizing max-min normalization and one-hot encoding. 

(3) \textbf{Model Design:} Instead of considering end-to-end learning, we design a two-stage learning method to decouple feature embedding and uncertainty label learning. In the first stage, we design an efficient query embedding model to achieve transferable embedding vectors. In the second stage, we design a lightweight model to predict knob uncertainty labels. The model inputs the query embedding vectors and knob configurations to estimate the uncertainty distribution label. Compared to an end-to-end learning model, our decouple learning design enables sufficient embedding training to obtain high-quality embedding features and efficient model transfer by fine-tuning the lightweight prediction model.

The main contributions of this paper are as follows:

     (1) We propose a novel concept of the uncertainty-aware knob configuration estimation and define the problem of uncertain estimation in Section~\ref{sec:background} to enhance the knob tuning process. 
     
    (2) We design transferable feature representation for the uncertainty estimation task, including the transferable query representation learning method in Section~\ref{sec:embedding} and the transferable knob configuration encoding method in Section~\ref{sec:knobencoding}.

    (3) We propose a query uncertainty-aware knob classifier, \textsf{KnobCF}, and introduce the enhanced knob tuning algorithm in Section~\ref{sec:tuning}. Our method could effectively reduce the time consumption of knob tuning evaluations while maintaining the knob tuning results.
    
    (4) We conduct experiments in Section~\ref{sec:eval} on four open-source benchmarks, demonstrating that our method effectively reduces useless evaluations and improves the tuning results.

\section{Overview}
\label{sec:background}
In this section, we introduce the background knowledge of the knob tuning problem, formalize the uncertainty-aware knob estimation, and present the overview of our method. 

\subsection{Preliminary}
\noindent\textbf{Knob Tuning Process.} As shown in Figure~\ref{fig:tuning}, we present the typical process of knob tuning, and the main components are shown as follows:

(1) The knob space defined as $K = \{K_1,..., K_n\}$ could be identified based on a certain database environment and workload. Generally, we could obtain the adjustable knob space according to resource limitations and DBMS types. Then, existing methods utilize knob selection methods~\cite{kanellis2020too} to reduce knob space for certain workloads. The knob tuning aims to find the optimal knob configuration from the filtered knob space. 

(2) For the search model, we could roughly divide the existing tuning model into three types: the rule-based heuristic search model~\cite{Bestconfig}, the RL-based search model~\cite{zhang2019end}, and the BO-based search model~\cite{GPTuner}. Existing methods typically start the tuning process by sampling several knob points to initialize the search model. Then, the model is responsible for recommending diverse knob configurations to further find the optimal solution.

(3) The recommended knob configuration could be generated by several principles. Specifically, RL-based models utilize the knob configuration with the highest potential reward as the next recommended knob configuration. BO-based models employ the knob configuration with the maximum acquisition function value. And the rule-based models utilize the heuristic greedy rules to select the knob configuration.

(4) Existing knob tuning methods typically utilize one-time or multi-time evaluations to obtain the corresponding reward for a certain knob configuration. Different from these methods, we consider the uncertainty-aware estimation for knob tuning, to improve the evaluation efficiency of knob tuning.

\noindent\textbf{Uncertain Estimation.} Naturally, we could formalize the uncertain estimation task for knob tuning as follows: given a knob configuration $k \in K$ and the workload $W$ (a set of queries), the uncertain distribution of workload execution time could be defined as $T(k, W) \sim f(\xi_{k,W})$, where $\xi_{k,W}$ is the random variable. The $f(\xi_{k,W})$ represents the uncertain distribution of workload execution time in knob configuration $k$. Considering the complexity of directly modeling workload, we could simplify the uncertain estimation task as follows: $T(W, k) = \sum_{q_i \in W} T(k, q_i) $, where $T(k, q_i)  = f(\xi_{k,q_i})$ is the uncertain distribution of query $q_i$ in knob configuration $k$. 

\subsection{Problem Definition}
In this paper, we aim to design an uncertain-aware knob estimator to enhance the knob tuning task. Specifically, given the knob configuration $k$ and the query $q$, our goal is to learn the uncertain distribution of query execution time defined as $f(\xi_{k,q})$. Generally speaking, we could formalize the uncertain-aware knob estimation task as a regression task. The input of the regression task is the knob configuration $k$ and the query $q$, and the output is the uncertain distribution range of the query execution time.

However, the above regression task requires multiple evaluations for each query and each knob configuration to construct the train data, which is time-consuming in the database management system. To optimize this problem, we consider two novel designs for the learning goal. 

(1) We define the uncertain prediction task as a classification task. In fact, the category information of the knob configurations is sufficient for detecting the useless evaluations. For a new configuration, we only judge the uncertainty category instead of predicting the exact uncertainty distribution. If we observe that similar knobs have been evaluated before, we can directly use the historical evaluations to predict the current distribution. 

(2) The second design is to improve the single uncertain distribution to the joint uncertain distribution. As we discussed in Section~\ref{sec:intro}, uncertainty distributions among different knobs can complement each other to save time consumption of query evaluation. Thus, we further improve the uncertain estimation task to a joint estimation. Given the knob space $K$ and the workload $W$ (a set of queries), the uncertain distribution of workload execution time could be defined as $T(K, W) = \sum_{q_i \in W} T(K, q_i) $, where $T(K, q_i) \sim f(\xi_{K,q_i})$ represents the uncertain distribution of query $q$ in knob space.  

Based on the above two novel designs, we define our classification-based knob estimator as follows. \textsf{Input:} The knob configuration $k \in K$ and the query $q \in W$. \textsf{Output:} The category label of uncertain distribution $f(\xi_{k,q})$. Furthermore, we introduce the construction of our joint uncertain distribution estimator and the detailed utilization of our knob estimator in Section~\ref{sec:feature}~\ref{sec:tuning}.

\textbf{Example 1:} We present an example to illustrate the above definition. Given the knob space $K = \{k_1, k_2, k_3, k_4, k_5\}$ and the workload $W = \{q_1\}$. Assume that $f(\xi_{K,q_1}) = \sum_{i = 1}^{2} \pi_i * \mathcal{N} (\mu_i, \sigma_i) = \pi_1 \times \mathcal{N} (\mu_1, \sigma_1^2) + \pi_2 \times \mathcal{N} (\mu_2, \sigma_2^2)$ the joint uncertain distribution of query $q_1$. We observe this joint distribution consists of two Gaussian distributions, $\mathcal{N} (\mu_1, \sigma_1^2)$ and $\mathcal{N} (\mu_2, \sigma_2^2)$. Thus, for each $k_i \in K$, knob configuration has three potential category label, $[1,0]$ (the uncertain distribution of $k_i$ is involved in $\mathcal{N} (\mu_1, \sigma_1^2)$.), $[0,1]$, $[1,1]$(the uncertain distribution of $k_i$ is combined by $\mathcal{N} (\mu_1, \sigma_1^2)$ and $\mathcal{N} (\mu_2, \sigma_2^2)$.).

\subsection{Solution Overview}
Knob tuning is a time-consuming task that involves multiple evaluations to determine the optimal configuration in a high-dimensional space of knobs. The efficiency of these evaluations directly impacts the overall efficiency of knob tuning. 
A natural approach is to leverage the observation that if there are useless evaluations, similar performance distributions exist among knobs. We exploit this property by designing a knob uncertainty classifier to group similar knob configurations into the same distribution. This design allows us to estimate new knob configurations based on existing evaluations without individually evaluating each one. 

\begin{figure}[ht]
    \centering
    \includegraphics[width=\linewidth]{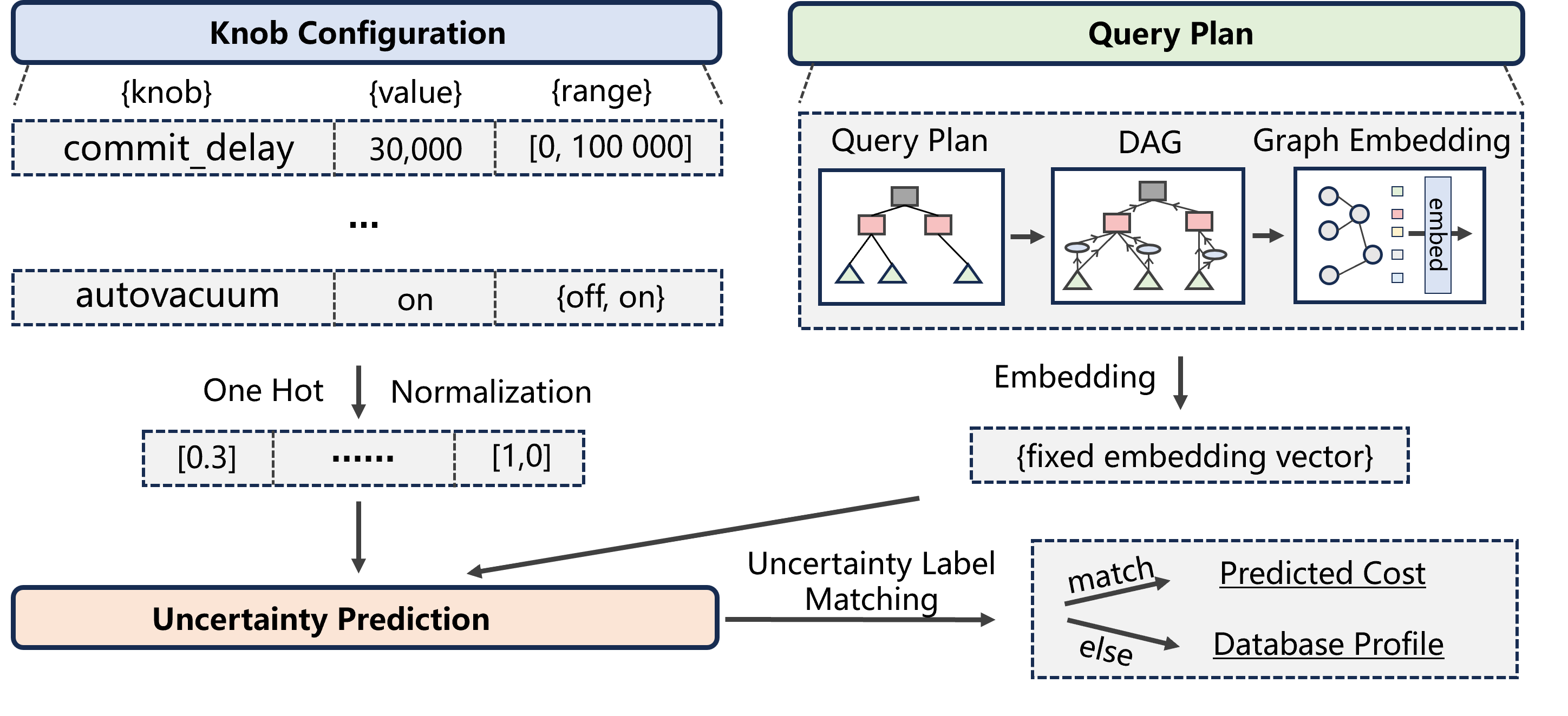}
    \caption{The Workflow Of Our Uncertainty-Aware Knob Classifier.}
    \label{fig:overview}
\end{figure}

Specifically, we show the workflow of our method in Figure~\ref{fig:overview}, consisting of two fundamental components: (1) a transferable feature embedding model that encodes the query plan and knob importance into an embedding vector, which is used to obtain a high-quality embedding representation for various queries and will be discussed in Section~\ref{sec:feature}; (2) an uncertainty-aware knob classifier that predicts the distribution category for a given knob configuration and query embedding vector, and will be discussed in Section~\ref{sec:tuning}. Then, if KnobCF matches similar historical evaluations, our model directly returns an estimated cost for the current query instead of practical query execution in the database. 

\begin{figure*}[t]
    \centering
    \includegraphics[width=0.85\linewidth]{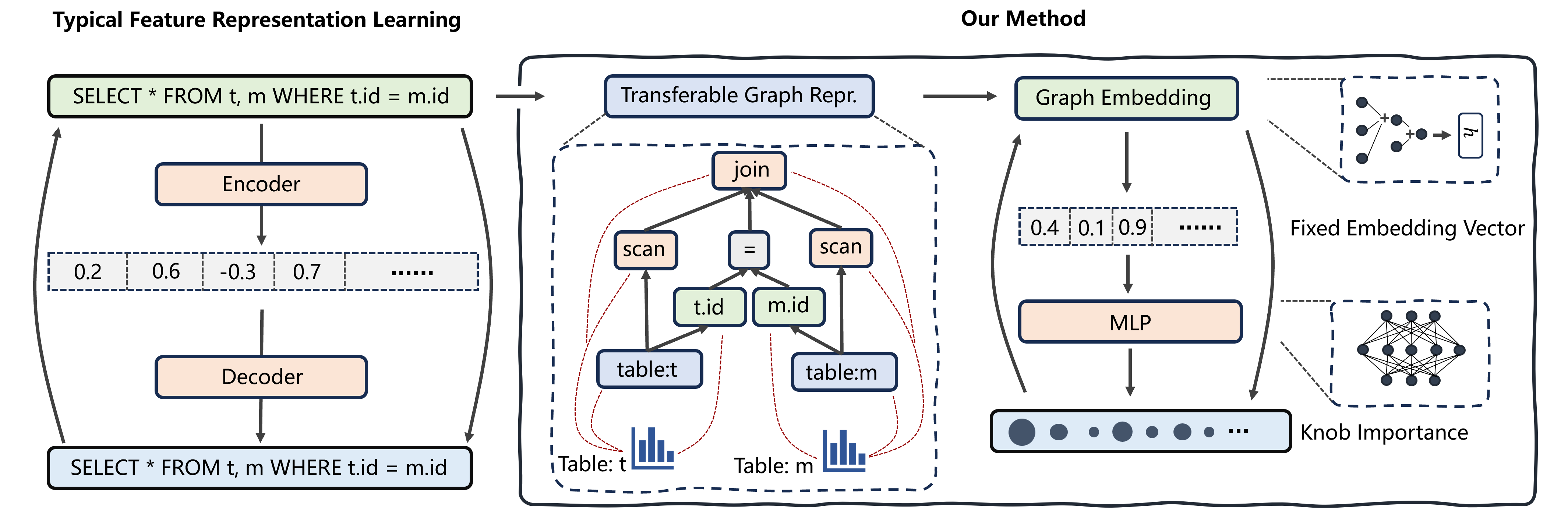}
    \caption{The architecture of query representation learning model.}
    \label{fig:query}
\end{figure*}

To further explain the workflow of our uncertainty-aware knob classifier, we present an example based on Example 1. Given the category labels $L = \{[1,0], [0,1], [1,1], [1,0], [0,1]\}$ for all knob configurations $K$ under $q_1$, the label $[1,0]$ corresponds to the uncertain distribution $\mathcal{N} (\mu_1, \sigma_1)$, $[0,1]$ corresponds to $\mathcal{N} (\mu_2, \sigma_2)$, and $[1,1]$ corresponds to both $\mathcal{N} (\mu_1, \sigma_1)$ and $\mathcal{N} (\mu_2, \sigma_2)$. For a new knob configuration $k_6$ and query $q_1$, we first encode the query plan and knob configuration into embedding vectors. Then, we input the embedding vectors into the uncertainty-aware knob classifier to predict the category label $[1,1]$. Then, we match the historical labels to determine whether to evaluate the knob configuration. In this case, $k_6$ has the same label as $k_1$ and $k_4$. Therefore, we can estimate the performance of $k_6$ based on the evaluations of $k_1$ and $k_4$ without conducting a new evaluation.

\section{Transferable Feature Representation Learning}\label{sec:feature}
Feature representation is the fundamental task for our uncertainty-aware knob classifier, directly influencing the efficiency of model training, inference, and transfer. However, it is challenging to process the diverse queries and the high-dimensional knobs to obtain the high-quality feature representation. In this section, we introduce our transferable query feature embedding model in Section~\ref{sec:embedding} and the transferable knob configuration encoding in Section~\ref{sec:knobencoding}.

\subsection{Transferable Query Feature Embedding}
\label{sec:embedding}
Generally speaking, existing AI-driven methods~\cite{QueryFormer, end2end-ce, hilprecht2022zero} have proposed some effective feature representation for queries. For example, the QueryFormer~\cite{QueryFormer} proposed a tree-transformer model to learn embedding representations for queries. Hilprecht et al.~\cite{hilprecht2022zero} proposed a zero-shot cost model for query plan encoding which focuses on the transferable feature representation learning. However, these estimators focus on query optimization and index tuning, lacking the consideration of knob tuning. Moreover, the uncertainty prediction of knob tuning faces a large knob candidate space, making it difficult to model the uncertainty distribution. To address this challenge, we design a downstream task-related query embedding model to provide high-quality embedding vectors for knob uncertainty classification.

Specifically, we combine the knob characteristics with the query embedding representation. Our key idea is to combine the knob importance~\cite{kanellis2020too, sullivan2004using, tan2019ibtune} with the query embedding representation, which could benefit the uncertainty classification task. Specifically, the knob importance calculated by SHAP~\cite{lundberg2017unified} is typically utilized to filter useless knobs in existing works~\cite{zhang2022facilitating}. The important information identifies the knob sensitivity for queries, largely influencing the knob uncertainty category. Also, similar queries have similar sensitivity to the same knobs. The knob importance information could be directly related to queries. For efficiently obtaining the knob importance information, we utilize popular openFE~\cite{zhang2023openfe} to calculate the knob importance for each query. Compared to SHAP~\cite{lundberg2017unified}, OpenFE~\cite{zhang2023openfe} utilizes the lightGBM to efficiently calculate the dimension importance, bringing fast and accurate knob importance score. Furthermore, the knob importance could be manually adjusted by the database administrator to better suit the practical environment. 

With the knob importance obtained, we combine the knob importance with the query embedding representation. As shown in Figure~\ref{fig:query}, we replace the output query plan with the knob importance for query representation learning. We make this revision due to two considerations. On the one hand, some machine learning methods~\cite{QueryFormer, donoso2023astromer} have clarified that the downstream task-related embedding method could achieve model learning more efficiently than the typical encoder-decoder model, which is a general schema of representation learning. On the other hand, compared to the uncertainty distribution, it requires a small part of knob-performance training data to calculate the knob importance for a certain query. In the model transfer, we could implement efficient retraining to adapt to workload drift. 

Specifically, our query representation learning model consists of three main components: 

(1) Before implementing the representation learning, we encode the original query plan to the vector representation. Considering the model transfer, we adopt the transferable feature representation~\cite{hilprecht2022zero} and construct the directed acyclic graph for queries. As shown in Figure~\ref{fig:query}, a query plan is represented by a directed acyclic graph consisting of operator nodes, predicate nodes, table nodes, etc. Then, each node consists of the transferable representations. For example, the column feature consists of the column type and statistical characteristic instead of column-length one-hot codings. 

(2) We employ the bottom-up graph convolution embedding method~\cite{gilmer2017neural} to process the plan graph, which could be flexibly extended to process diverse plan graphs. Due to the graph convolution design, our query embedding model could process different query plan structures. Then, we regard the hidden output of the query plan root node as the query representation learning results, i.e. the fixed embedding vector. This fixed embedding vector will be used to predict the knob uncertainty category in Section~\ref{sec:tuning}.

(3) We design a knob importance leaner based on a three-layer neural network, which processes the embedding vector to knob importance. In our feature representation learning model, the knob importance information will be combined with the query embedding vector by the graph embedding parameters updated by backpropagation. In the training process, with limited time consumption, we could gather sufficient query-importance data to obtain a high-quality query embedding vector, enhancing the downstream knob uncertainty prediction model.

\subsection{Knob Configuration Encoding}\label{sec:knobencoding}
In this section, we design a transferable knob configuration encoding method which is the basis for model transfer. Generally speaking, different database management systems (even different versions of the same DBMS) have different kinds of knobs, which makes it incredible to design a universal knob encoding method. 
Thus, we limit our focus on the same DBMS version to design a transferable knob encoding method. Specifically, the database knobs could be divided into two categories: numerical knobs and non-numerical knobs. 

Actually, it is easy to obtain the universal encoding for numerical knobs because the same DBMS version has database knobs. We only need to normalize the knobs to process the varying scale of different tuning tasks. For numerical knobs, we directly utilize the max-min normalization as shown in Formula~\ref{eqn:normalization} to normalize the different scales of potential knob spaces. 
\begin{equation}
    \label{eqn:normalization}
    x' = \frac{x - x_{min}}{x_{max} - x_{min}}
\end{equation}

Where $x$ is the original knob value, $x_{min}$ and $x_{max}$ are the minimum and maximum values of the knob space. After normalization, the numerical knob could be encoded as a fixed-scale vector. For non-numerical knobs, we utilize the one-hot encoding to obtain the universal encoding. For example, the `autovacuum' knob of PostgreSQL 14 has two potential values: `on', `off'. Thus, we encode the `autovacuum' knob as $[1,0]$ or $[0,1]$.

After obtaining the encoding, we directly concatenate them to obtain the final knob configuration encoding. Furthermore, from the aspect of model transfer, there exists a contradiction between comprehensive knob encoding and filtered knob encoding. On the one hand, we could encode all the potential knobs of a certain DBMS version to support the various tuning tasks. While comprehensive knob encoding may bring high input dimension, leading to the underfitting problem and poor model prediction~\cite{zhang2019overfitting}. On the other hand, the filtered knob encoding could be obtained by filtering the useless knobs, which could largely reduce the knob search space and is the most important preprocessing of existing tuning tasks. However, the filtered knob encoding may bring limited knob representation, resulting in the overfitting problem and poor model transfer~\cite{han2019learning}. 

 To balance the above contradictions, we simply utilize the union knob set of the important knobs in the historical tuning task. This problem could be further improved by some dimensionality reduction methods like the random projection in llamatune~\cite{kanellis2022llamatune}. And these methods could be easily integrated into our knob encoding method.

\section{Uncertainty-aware Knob Tuning}
\label{sec:tuning}
Based on the high-quality feature representation, we aim to predict the category label of a certain knob configuration. With similar historical knob configurations, our method could directly predict the latency category of the knob configurations to enhance the knob tuning. Then, we introduce the uncertainty-aware knob classifier in Section~\ref{sec:uncertainty-estimation} and the specific uncertainty-aware knob tuning method in Section~\ref{sec:knob-tuning}.

\subsection{Uncertainty-aware Knob Classifier}\label{sec:uncertainty-estimation}

In this section, we introduce the uncertainty-aware knob classifier, which could be used to predict the query latency distribution. Basically, we concatenate the query embedding vector and the knob configuration vector to form the input of the classifier. Taking advantage of our transferable encoding, the classifier could be directly applied to various tuning tasks on the same DBMS version.

\noindent \textbf{Gaussian category label} After obtaining the classifier input, we consider designing the output label. Specifically, the output label design of our classifier faces two challenges.

\textbf{I: how to obtain the training label of the classifier?} A fundamental problem is Which configurations should be grouped into one category. A straightforward method is to make multiple evaluations for each knob configuration and divide the knob configurations according to some statistics, such as mean and variance. However, this method is time-consuming and not practical in the real-world tuning task. 

\textbf{II: how many categories should we design?} The number of categories is a key factor that affects the performance of the classifier. On the one hand, the small number of categories may not be able to distinguish the differences among different knob configurations, leading to the different knob configurations being divided into the same category. On the other hand, if we design a large number of categories, the classifier may underfit the training data and result in a low prediction hit rate.

To address the above challenges, we design the Gaussian category labels based on the mixed Gaussian model (GMM)~\cite{reynolds2009gaussian}. Specifically, we assume that the joint query latency distribution of knob configurations follows the mixed Gaussian distribution. We make this assumption based on two aspects. (1) \textbf{Knob Tuning Aspect:} Multiple existing knob tuning methods~\cite{thummala2010ituned, van2017automatic, kunjir2020black} design the Gaussian-based surrogate model to fit the relationships between the knob configuration and the corresponding performance. This indicates that the performance of knob configurations may follow Gaussian distribution. (2) \textbf{Query Estimation Aspect:} Existing methods~\cite{wu2014uncertainty, hartwig2019approximate} make similar assumptions to predict cardinality and query latency. This clarifies that the performance of the majority of queries satisfies Gaussian distribution in practical database management systems.

\begin{figure}[t]
    \centering
    \includegraphics[width=1\linewidth]{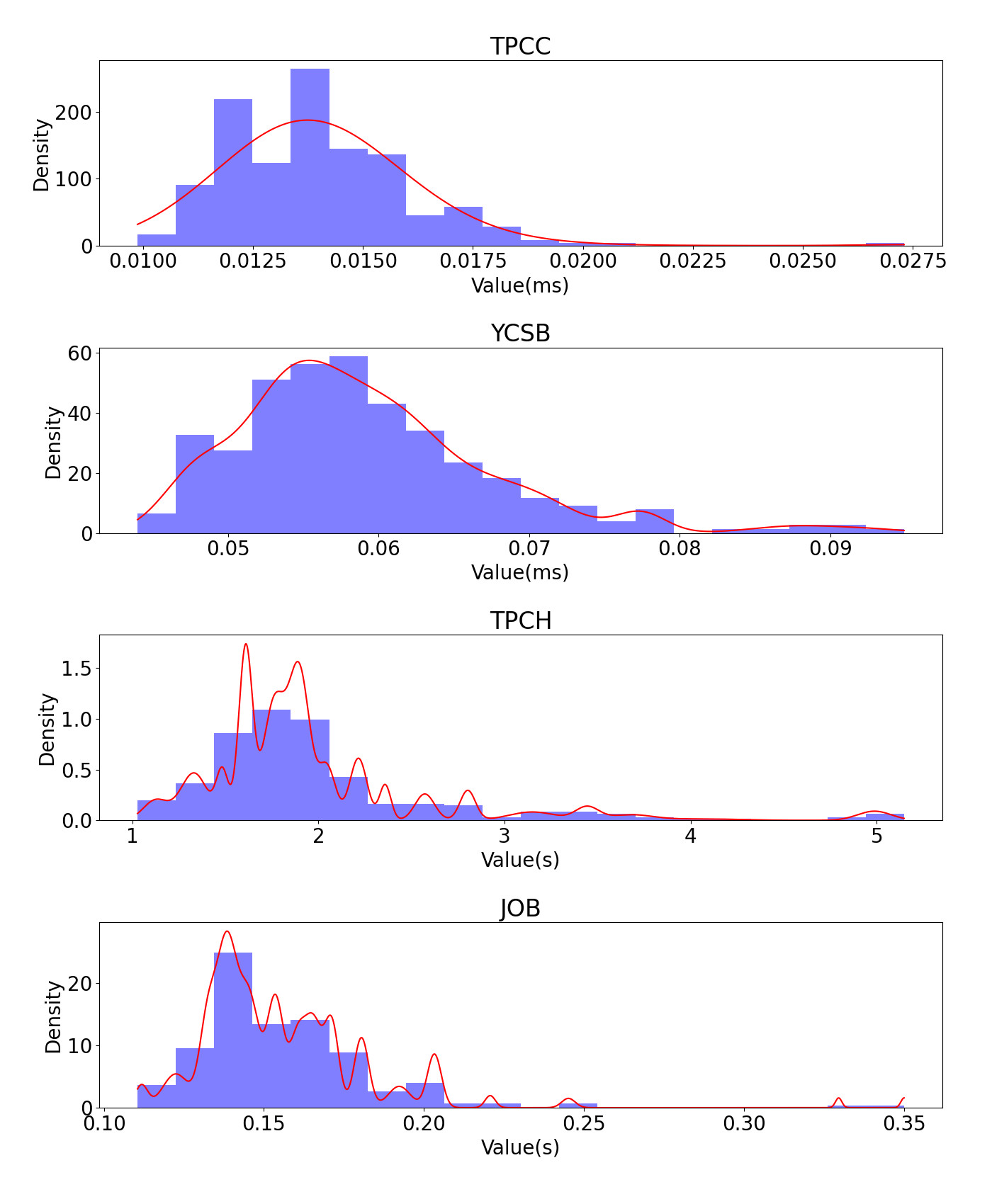}
    \caption{The mixed Gauss distribution of different queries under 300 knob configurations.}
    \label{fig:queryexa}
\end{figure}

Thus, we think the mixed Gaussian is the best choice for the joint latency distribution of knob configurations. Then, we can address the first challenge. We utilize the mixed Gaussian model to analyze the mixed Gaussian distribution for knob configurations. Then, we could obtain the category label for every knob configuration according to the clustering results of GMM. As shown in Figure~\ref{fig:queryexa}, we plot the uncertainty distribution of different queries under 300 knob configurations. We observe that these points could be represented by a mixed Gaussian distribution. Once we evaluate partial points, we could predict other knob configurations based on the mixed Gaussian distribution. 

Based on the GMM design, we address the second challenge. In our classifier model, we could utilize a small number of output bits to represent multiple categories. On the one hand, the number of categories increases exponentially with the parameters of the model, i.e. $|C| = 2^n$ where $n$ is the number of mixture components in GMM, and $|C|$ is the number of categories. For example, we set the number of output bits is 3. Then, we represent the eight category labels by 3 bits, i.e., [0,0,0], [0,0,1] ... [1,1,1]. Also, from the example in Figure~\ref{fig:queryexa}, we observe that although we make evaluations for 300 knob configurations, there exist limited categories for every query. Thus, the limited output bits are sufficient to represent the uncertainty distribution of knob configurations.

\noindent \textbf{Knob Classifier} Based on the label, we propose the architecture of the uncertainty-aware knob classifier. Its input is the concatenated vector of the query embedding and the knob configuration. Its output is the binary category label of the knob configuration. 

The specific classifier consists of three fully connected layers with the LeakyReLU activation function~\cite{radford2015unsupervised}. We utilize the cross-entropy loss function~\cite{zhang2018generalized} to optimize the model for the training process. Actually, our classifier design is a simple neural network, which is easy to train and transfer to different tuning tasks. This simple structure benefits from our simple classification problem definition and feature representation learning. These preprocessing allow our classifier to focus on uncertainty classification based on high-quality embedding vectors and knobs, greatly reducing the learning burden of the classification prediction model.

\subsection{Uncertainty-aware Knob Tuning}\label{sec:knob-tuning}
\begin{algorithm}[t]
    \caption{Uncertainty-aware Knob Tuning}
    \label{alg:tuning}
    \begin{algorithmic}[1]
        \Require workload $W$, the number tuning iteration $N$, the knob space $S$, the trained uncertainty-aware knob classifier $KnobCF$
        \State $L \gets \emptyset$ //Initialize the label set\footnote{note here}
        \State $P \gets LHS(S)$ Generate initial knob points using LHS sampling
        \State $D \gets \emptyset$ Initialize the knob-performance datasets
        \For{$p \in P$}
            \State $total_p \gets 0$ 
            \For{$q \in W$}
                \State $time = DBMS(p,q)$ Execute the query $q$ with the knob configuration $p$
                \State $label \gets KnobCF.predict(p,q)$ Predict the category label of the knob configuration $p$ and query $q$
                \State $total_p \gets total_p + time$ 
                \State $L.add(label,time)$ 
            \EndFor
            \State $D.add(p,total_p)$
        \EndFor
        \State $tunner.train(D)$ Initialize the knob tunning model
        \For{$i=1$ to $N$}
            \State $p \gets tunner.recommend()$ Recommend the next knob configuration
            \State $total_p \gets 0$ 
            \For{$q \in W$}
                \State $label = KnobCF.predict(p,q)$ 
                \If{$KnobCF.judge(label, L)$}
                    \State $time \gets KnobCF.estimate(L, Mean, label)$ Estimate the performance according to the historical evaluations and category Label
                \Else 
                    \State $time \gets DBMS(p,q)$ Execute the query $q$ with the knob configuration $p$
                    \State $L.add(label,time)$ 
                \EndIf
                \State $total_p \gets total_p + time$
            \EndFor
            \State $D.add(p,total_p)$
            \State $tunner.update(D)$ Update the knob tunning model
        \EndFor
    \Return $tunner.p*$
    \end{algorithmic}
\end{algorithm}

\noindent \textbf{Knob Tuning} For a tuning task, our knob classifier could be applied directly to attach labels for knob configurations. Then, the historical knob configurations with the same category could be utilized to predict the performance of the new knob configuration. In particular, in this prediction process, we only utilize historical evaluations of the same tuning task to calculate the estimated performance. Thus, even in model transfer scenarios, our method could keep the predicted performance within the scope of practical query distribution.

Typically, the knob tuning process consists of two main parts: the initialization part which is responsible for constructing the knob turner, and the iterative part which aims to iteratively find the optimal knob configuration. In this section, we integrate our knob classifier into the initialization process and iterative optimization process to enhance the knob tuning. Specifically, we introduce the uncertainty-aware knob tuning method as shown in Algorithm~\ref{alg:tuning}. The input of our algorithm consists of four parts, including the workload $W$, the number of tuning iterations $N$, the knob candidate space $S$, and the trained uncertainty-aware knob classifier $KnobCF$.

In the initialization part, we first generate the initial knob configurations using Latin Hypercube Sampling (LHS)~\cite{helton2003latin, stein1987large}. In existing work~\cite{kanellis2020too, kunjir2020black, thummala2010ituned, zhang2021restune}, the points $P$ are used to initialize the tuning model. Then Lines 4-11 complete the single-point evaluations of these sampling knob configurations. Importantly, even if we utilize single-point evaluations for knob configurations, our method still captures the joint uncertainty distribution by the multiple evaluations on the same category. Meanwhile, we predict the category label for each query in Line 8. And record the corresponding label and time information in Line 10, which will be used to estimate the future knob configurations. Then, Line 12 completes tuning model initialization based on the knob-performance datasets.

In the iterative knob tuning process, Line 14 recommends the next knob configuration based on the knob tuning model, such as the surrogate model of Bayesian Optimization~\cite{thummala2010ituned, van2017automatic, zhang2022towards} and the agent of reinforcement learning~\cite{cai2022hunter, li2019qtune, zhang2019end}. Then, Lines 16-23 evaluate the performance of the recommended knob configuration $p$. For each query $q$, Line 17 predicts the corresponding category labels by the trained uncertainty-aware knob classifier. If the category label has appeared in the historical evaluations, Line 19 directly estimates the query latency based on the mean value of historical evaluations. The mean value could represent the stable performance of a certain category. Otherwise, Line 21 executes the query with the recommended knob configuration, and Line 22 updates the label set. Then, Line 24 updates the knob-performance datasets and the knob tuning model, respectively. Finally, the algorithm repeats the iterative knob tuning process until the number of iterations is reached and returns the optimal knob configuration with minimal total latency.

Note that the uncertainty-aware knob tuning method is a general framework for knob tuning tasks, which could be easily integrated into existing knob tuning methods, such as GPTuner~\cite{GPTuner}, CDBTune~\cite{zhang2019end}, etc. For time consumption, our method only requires some extra model inference time, which could be accelerated by GPU and label matching time for limited categories, which is significantly lower than the query execution time. 

\noindent \textbf{Model Transfer} Furthermore, another important task is how to obtain a mature-trained knob classifier for tuning tasks. The basic idea is to gather a sufficient training set of the current tuning task to complete the knob classifier training. Although we could obtain a high-quality knob classifier for the current tuning task, obtaining sufficient training data is time-consuming and impractical for a real-world tuning task. Thus, we consider to obtain the trained knob classifier from historical tuning tasks. As we introduced in Section~\ref{sec:feature}, our method provides a transferable feature representation, including transferable query embedding and knob configuration representation. Thus, we could directly utilize the historical tuning tasks to train the knob classifier. Then, we could achieve the zero-shot model transfer and utilize the trained model on unseen tuning tasks.

Although the zero-shot model transfer is a good choice for saving time consumption, the zero-shot knob classifier still requires high-quality and representative historical tuning tasks to complete sufficient pretraining. If current tuning tasks have a large difference from historical tuning tasks, the performance of the zero-shot knob classifier may be degraded. In addition, the zero-shot knob classifier could not provide specialized estimation services for certain tuning tasks because the classifier is trained on multiple tuning tasks.

To address these problems, we consider a few-shot model transfer method instead of zero-shot one. Benefiting from our decoupled classifier model, we could utilize different transfer mechanisms for the feature representation learning model and the uncertainty prediction model. Firstly, feature representation learning is responsible for processing diverse queries and obtaining the high-quality embedding feature, which requires sufficient training data. Thus, we are not required to adjust the representation learning model for the current tuning task. Secondly, our uncertainty prediction model is lightweight enough to easily make few-shot retraining. We could utilize the initialization configurations of the evaluated knobs and the previous iterations of the knob tuning to fine-tune our uncertainty prediction model. Importantly, we do not implement extra evaluations for the few-shot retraining instead of only utilizing the evaluations of the tuning process, which has high time efficiency in practical tuning tasks.

We present an example to illustrate how to achieve few-shot training for KnobCF. Given two historical tuning tasks, we have evaluation observations as $O_1 = \{(k_1, q_1, c_1), (k_2, q_2, c_2), ...\}$ and $O_2 = \{(k_1, q_1, c_1), (k_2, q_2, c_2), ...\}$, where $k_i$ is the knob configuration, $q_i$ is the query, and $c_i$ is the corresponding query latency. We could utilize sufficient historical evaluations to complete the KnobCF pretraining, including training the feature representation learning model and the uncertainty prediction model. Then, for the current tuning task, we could obtain the evaluation observations of the initialization phase as $O_{init} = \{(k_1, q_1, c_1), (k_2, q_2, c_2), ...\}$. Also, we could obtain the evaluation observations of the first 30 iterations as $O_{30} = \{(k_1, q_1, c_1), (k_2, q_2, c_2), ...\}$. Then, we could utilize the $O_{init}$ and $O_{30}$ to finetune the uncertainty prediction model of KnobCF, improving the performance of the classifier on the current tuning task. In particular, our pretraining and finetuning process does not collect extra training data, which is efficient in practical tuning tasks.

\section{Experiment}\label{sec:eval}
In this section, we conduct extensive experiments to evaluate the effectiveness of our KnobCF, including the experiment settings in Section~\ref{sec:setting}, the evaluations of our uncertainty-aware knob classifier in Section~\ref{sec:classifier-eval}, and the evaluations of our uncertainty-aware knob tuning method in Section~\ref{sec:tuning-eval}, and the robustness analysis in Section~\ref{sec:robustness}.

\subsection{Experiment Setup}\label{sec:setting}

\subsubsection{Databases \& Workloads}

Our evaluation was conducted on four popular open-source benchmarks: \textsf{TPCH}, \textsf{JOB-light}, \textsf{YCSB}, and \textsf{TPCC}. In the \textsf{TPC-H} benchmark (scale factor = 10), we set the random seed to 1 and generate 22 instance queries based on 22 templates. Based on the IMDB dataset (size = 7049MB), this workload contains 70 queries generated by the job-light workload. For the \textsf{YCSB} benchmark, we generate 1000 queries for each of the three workloads: YCSB-a and YCSB-b, by utilizing `recordcount = 1000000`, `operationcount = 1000` settings. The YCSB-a workload has a read/write ratio of 50\%/50\%, the YCSB-b workload has a read/write ratio of 95\%/5\%. For the \textsf{TPCC} benchmark, with `warehouses=100` and `loadWorkers=4`, 256 queries were generated.

Since we make some assumptions for the query joint performance distribution (mixed Gaussian distribution) under different knob configurations, we conducted normality analysis on all queries across four benchmarks to check if queries follow a normal distribution. Then, we utilize the satisfied queries to make our evaluations and present the generated workloads in our github~\footnote{\url{https://github.com/AvatarTwi/KnobCF/benchmarks}}.

\subsubsection{Hardwares}
Our experiment is conducted in Windows 11, equipped with an Intel Core i7-12700H processor, 40GB of memory, 2.5TB of disk space, and an NVIDIA Geforce 3060Ti graphics card. Meanwhile, the Postgres database system version 14.4 is deployed on a Linux virtual machine with 4 cores, 16GB of memory, and 256GB of disk space.

\subsubsection{Implementation}
We implement our knob classifier in Python3 based on PyTorch-cuda~\cite{paszke2019pytorch} and the query graph embedding source code provided by Hilprecht et al~\cite{hilprecht2022zero}. The specific source code and benchmarks are publicly available at \url{https://github.com/AvatarTwi/KnobCF}.

\subsection{Evaluations of Uncertainty-aware Knob Classifier}\label{sec:classifier-eval}

In this section, we evaluate the uncertainty-aware knob classifier on extensive benchmarks, including the evaluation metrics in Section~\ref{sec:metrics}, the baselines in Section~\ref{sec:baselines}, and the experimental results in Section~\ref{sec:results}.

\subsubsection{baselines} \label{sec:baselines}
Our method is the first work to study the query uncertainty-aware knob configuration estimation. Thus, we could not directly compare our method with the existing query cost estimation methods. In our experiments, we replace our query embedding model with existing query encoding methods to evaluate the effectiveness of our method. Existing query plan encoding methods consist of three main types: (1) Tree-CNN, like Bao~\cite{Bao}. (2) Tree-RNN, like Plan-Cost~\cite{Tree-Rnn}. (3) Tree-Transformer, like QueryFormer~\cite{QueryFormer}. Among these, the QueryFormer proposes a general encoding method for various database tasks and outperforms other plan encoding methods~\cite{end2end-ce} in cost estimation tasks due to its attention mechanism. Thus, we choose the QueryFormer method as the baseline, called KnobCF(QueryFormer) to evaluate the effectiveness of our knob classifier.

\subsubsection{Evaluation Metrics} \label{sec:metrics}
We evaluate our uncertainty-aware knob classifier based on the following popular metrics, which are also widely used in existing works~\cite{liu2014strategy, vujovic2021classification, hossin2015review}. 

\textbf{Accuracy.} Accuracy shown in Formula~\ref{equ:accuracy} refers to the proportion of correctly predicted knob configurations among all knob configurations. It indicates the classifier's overall prediction performance. 

\begin{equation}\label{equ:accuracy}
accuracy=\frac{TP+TN}{TP+TN+FP+FN}.
\end{equation}

\textbf{Precision.} Precision shown in Formula~\ref{equ:precision} refers to the proportion of correct matched knob configurations among all matched knob configurations. It indicates the classifier's ability to predict the matched knob configurations correctly. 

\begin{equation}\label{equ:precision}
precision=\frac{TP}{TP+FP}.
\end{equation}

\textbf{Recall.} Recall shown in Formula~\ref{equ:recall} refers to the proportion of correct matched knob configurations among all actually matched knob configurations. It indicates the classifier's ability to capture actual matched knob configurations. 

\begin{equation}\label{equ:recall}
recall=\frac{TP}{TP+FN}.
\end{equation}

\textbf{Time Consumption.} Our training data comes entirely from evaluations of the tuning tasks. Thus, we ignore the data collection time and only utilize the model training time to clarify the time efficiency of our model preparation. 

\textbf{Inference Efficiency.} Inference efficiency refers to the time taken to predict the category label of a certain query plan. It is an important metric for the practical application of our uncertainty-aware knob classifier. We utilize the inference throughput per second to clarify the inference efficiency of our model.

\subsubsection{Experimental Results} \label{sec:results}
In this section, we evaluate the effectiveness of our uncertainty-aware knob classifier based on the above metrics. Since the QueryFormer utilizes the original features of the query plan and could not achieve model transfer, we make this comparison in static workload, i.e., we collect labeled data for each workload based on their tuning tasks and utilize 80\% as training data, 20\% as test data. 

\begin{table}[!t]
    \centering
    \caption{The efficiency of KnobCF in Extensive Benchmarks.} 
    \label{tab:query-plan-exp}
    \scalebox{0.85}{
  \begin{tabular}{ccccccc}
  \hline
  Type      & Metric    & TPCC  & YCSB-a & YCSB-b & JOB   & TPCH  \\ \hline
  \multirow{4}{*}{KnobCF}       & precision & 0.927 & 0.963  & 0.898  & 0.893 & 0.877 \\
  & recall    & 0.849 & 0.875  & 0.801  & 0.84  & 0.691 \\
  & accuracy  & 0.963 & 0.969  & 0.989  & 0.812 & 0.872 \\
  & time      & 51    & 160    & 97     & 18    & 18    \\
  \hline
  \multirow{4}{*}{\begin{tabular}[c]{@{}c@{}}KnobCF\\ (QueryFormer)\end{tabular}} & precision & 0.969 & 0.966  & 0.974  & 0.907 & 0.893 \\
  & recall    & 0.683 & 0.771  & 0.696  & 0.818 & 0.724 \\
  & accuracy  & 0.948 & 0.944  & 0.989  & 0.84  & 0.883 \\
  & time      & 41    & 230    & 52     & 18    & 16    \\ \hline
  \end{tabular}
  }
\end{table}

\noindent \textbf{Accuracy, Precision \& Recall.} Table~\ref{tab:query-plan-exp} shows the prediction accuracy, precision, and recall of our KnobCF and the KnobCF(QueryFormer) on the TPCC, YCSB, JOB-light, and TPCH workloads. In general, our KnobCF achieves an average accuracy of 0.921, an average precision of 0.911, and an average recall of 0.811 while the KnobCF(QueryFormer) achieves 0.920, 0.941, and 0.738 respectively. Specifically, the KnobCF method slightly outperforms the KnobCF(QueryFormer) method with an accuracy of 0.963 for TPCC and 0.969 for YCSB, compared to 0.948 and 0.944 for the Query Former method. For JOB-light and TPCH, we also achieve competitive accuracy of 0.812 and 0.872, compared to 0.840 and 0.883 for the QueryFormer method. This is because QueryFormer utilizes the original features of the query plan and designs a self-attention mechanism that captures more useful information to learn the long path relationships in query plans. Our KnobCF utilizes graph embedding to process the diverse query structure which focuses on learning the node relationships. Overall, even with the transferable feature representation, our KnobCF achieves a competitive prediction accuracy, precision, and recall compared to the KnobCF(QueryFormer).

\noindent \textbf{Time Consumption.} As shown in Table~\ref{tab:query-plan-exp}, the average training time of our KnobCF is 68.8 seconds, which is similar to the training time of 71.4 seconds of the KnobCF(QueryFormer). Especially, in TPCH which only has 22 queries, our KnobCF completes model training within 18 seconds. Actually, due to excluding the time consumption of training data collection, the training is directly determined by the GPU and the scale of the training set. Our KnobCF could achieve a second-level training time, which brings high time efficiency for practical tuning tasks.

\begin{figure}[htb!]
    \centering
    \includegraphics[width=\linewidth]{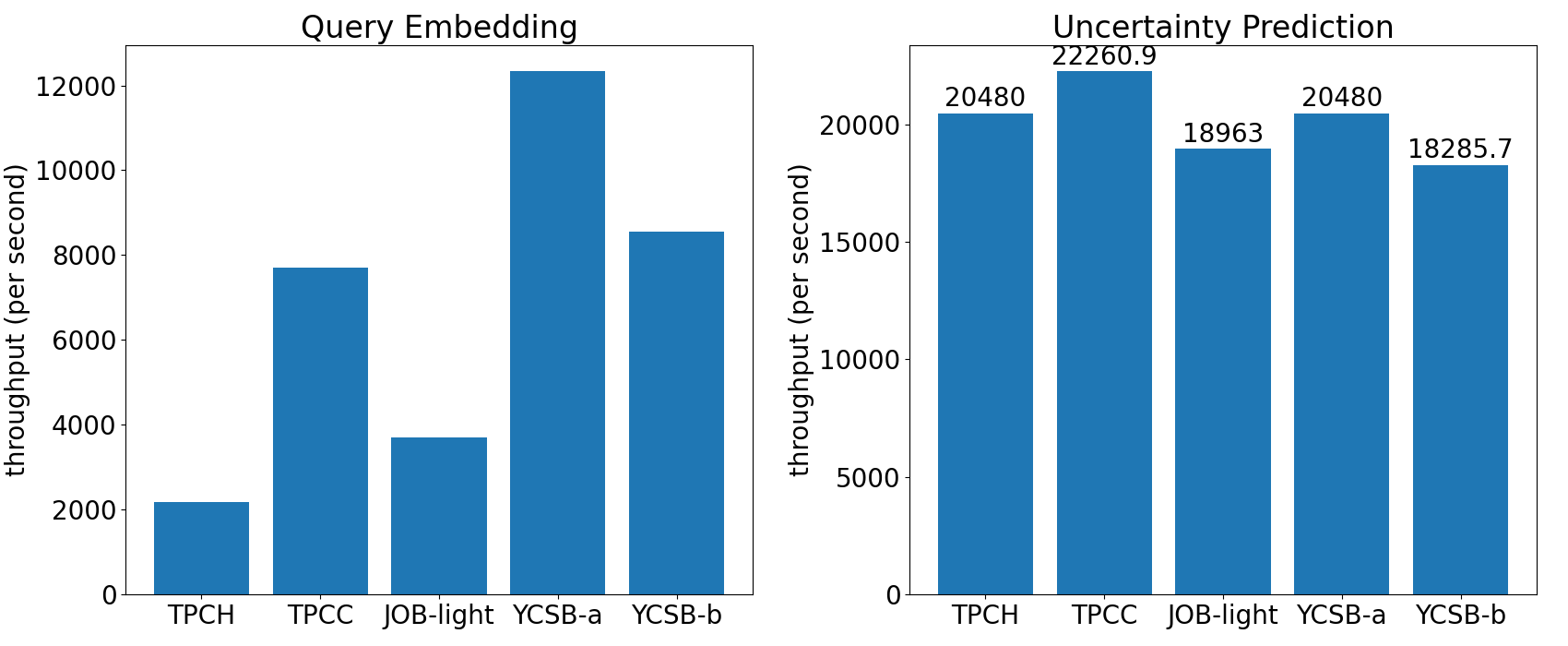}
    \caption{The inference throughput of our knob classifier.}
    \label{fig:inferencecf}
    \end{figure}
    
\noindent \textbf{Inference Efficiency.} As depicted in Figure~\ref{fig:inferencecf}, we present the inference throughput of our KnobCF, including the query embedding inference and uncertainty category inference. We observe that the query embedding inference throughput is directly related to the query structure complexity. Specifically, KnobCF reaches an inference throughput of 12331 queries per second on the YCSB-a workload while only 2174 queries per second on the TPCH workload. However, the query embedding vector could be reused for different knob configurations, which greatly reduces the query embedding time consumption. Meanwhile, the uncertainty category inference throughput is stable to extensive workloads and achieves an average inference throughput of 20093. This is because the uncertainty prediction model is a simple neural network model. This simple structure benefits from our uncertainty category definition and the high-quality representation learning, greatly improving the inference efficiency.

\subsection{Evaluations of Uncertainty-aware Knob Tuning}\label{sec:tuning-eval}
In this section, we evaluate the effectiveness of our uncertainty-aware knob tuning method, including the evaluation metrics in Section~\ref{sec:metrics2}, the baselines in Section~\ref{sec:baselines2}, and the experimental results in Section~\ref{sec:results2}.
\subsubsection{baselines} \label{sec:baselines2}
In our experiments, we utilize two state-of-the-art knob tuning optimizers, the Bayesian Optimization (BO)~\cite{thummala2010ituned} and the Deep Deterministic Policy Gradient (DDPG)~\cite{end2end-ce}. According to various works~\cite{van2017automatic, kunjir2020black, kanellis2022llamatune}, these optimizers are proven to work well in most situations. For BO optimizer, we utilize the SMAC3~\cite{lindauer2022smac3} implemented by Lindauer et.al. For the DDPG optimizer, we utilize the implementation provided by llamatune~\cite{kanellis2022llamatune}, which is the original neural network architecture used in CDBTune~\cite{zhang2019end}. Further, we integrate our KnobCF into the classical knob tuning optimizer, DDPG, to evaluate the effectiveness of our method. We design two kinds of implementations: the static uncertainty-aware knob tuning called DDPG(KnobCF) and the few-shot transfer knob tuning, called DDPG(few-shot). For static situations, we collect the 300 evaluations of the current tuning task to train our knob classifier. For the few-shot situation, we utilize the evaluations of the historical tuning tasks to complete pretraining and utilize the former 30 Iterations evaluations of the current tuning task to finetune our knob classifier. 

\begin{figure}[ht]
    \centering
    \subfigure{
		\includegraphics[width=\linewidth]{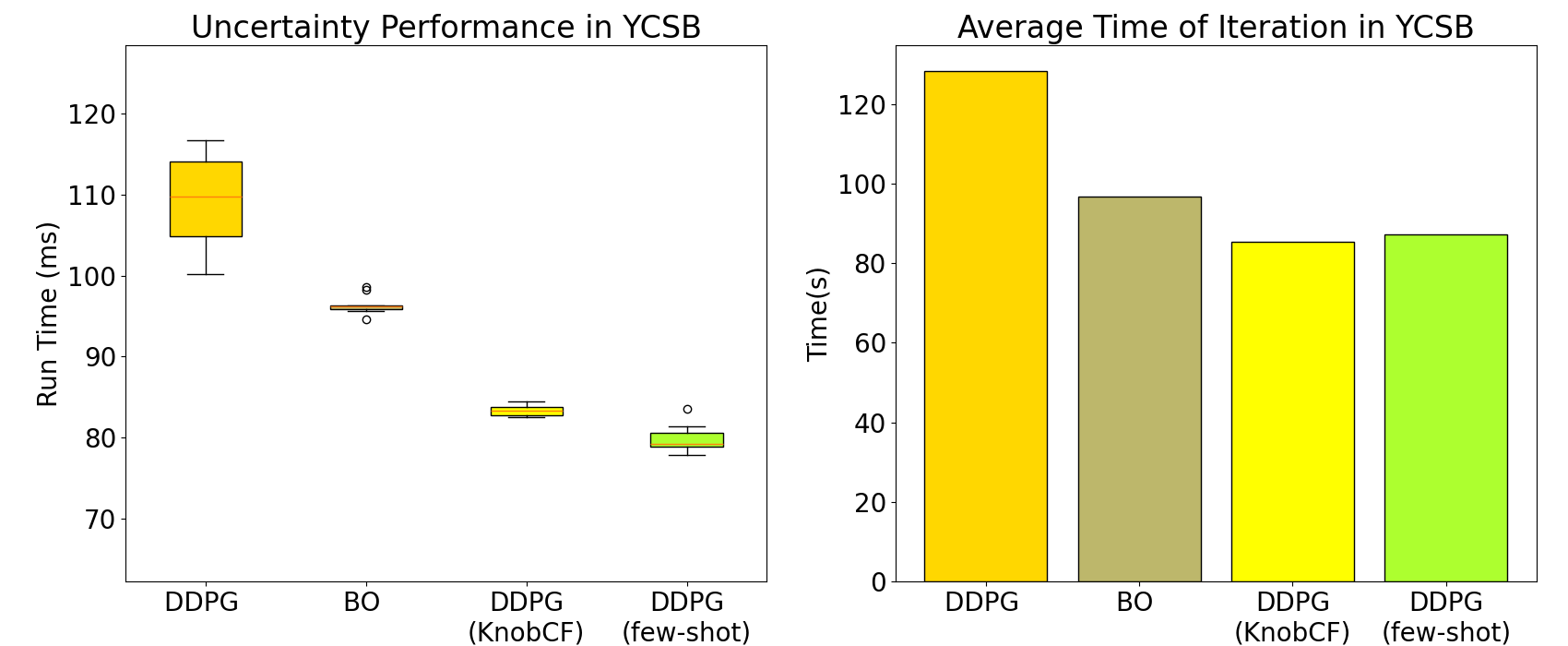}\label{fig:ycsb-bar} 
	}
    \subfigure{
		\includegraphics[width=\linewidth]{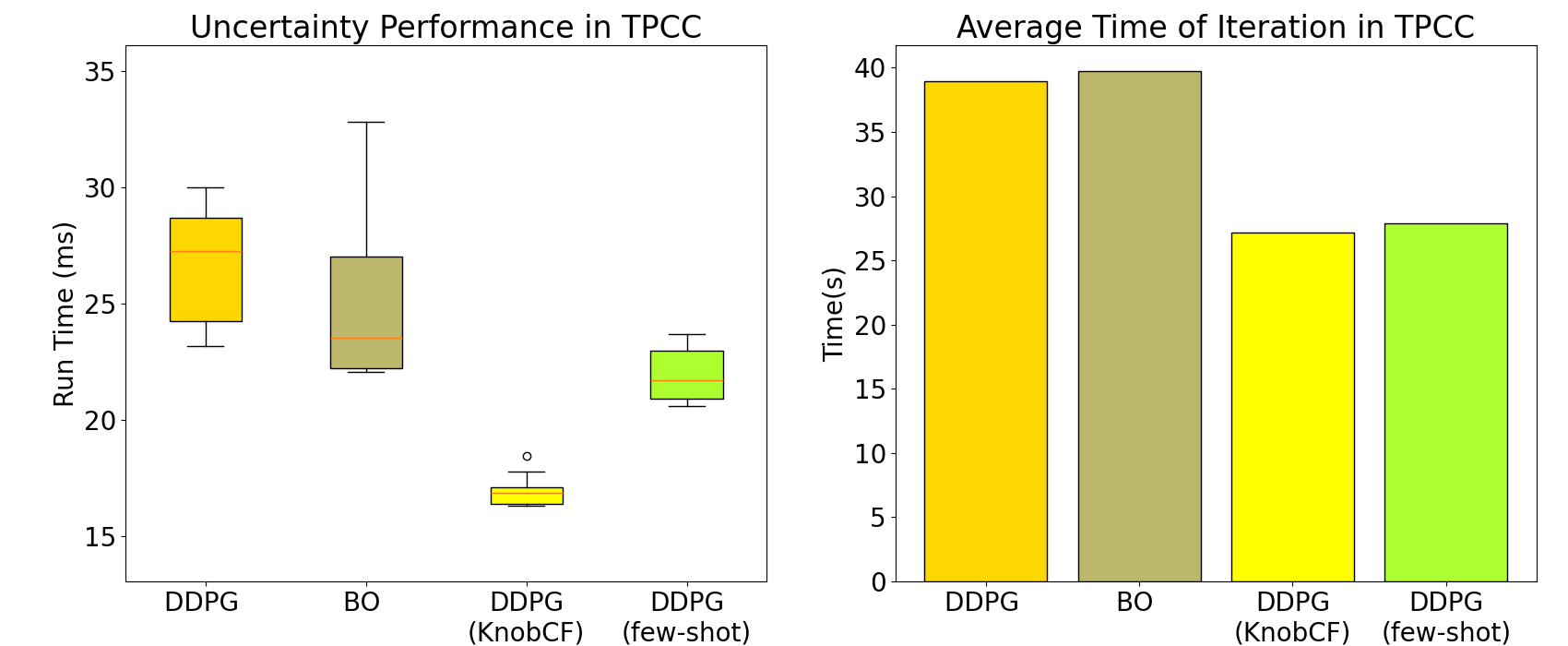}\label{fig:tpcc-bar}
	}
    \caption{The workload 90-th percentile tail latency distribution and the average time per iteration of different knob tuning methods.}
    \label{fig:tuningbar}
\end{figure}

\subsubsection{Evaluation Metrics} \label{sec:metrics2}
We utilize three typical metrics, which are also widely used in existing works~\cite{kanellis2020too, wang2022industry, kanellis2024nautilus}, to evaluate the effectiveness of our uncertainty-aware knob tuning method.

\noindent\textbf{DBMS Throughput}: This metric refers to the number of queries that could be executed per second. 

\noindent\textbf{The Uncertainty Latency}: We execute the full workload 10 times on the optimal knob configurations and utilize the 90-th percentile tail latency distribution as the uncertainty performance of workload.

\noindent\textbf{Average Time per Iteration}: This metric shown in Formula~\ref{equ:average} refers to the average time of an iteration of knob tuning, where $T_i$ is the time consumption of the i-th iteration. It reflects the time efficiency of our uncertainty-aware knob tuning.

\begin{equation}\label{equ:average}
    T =\frac{1}{n} \sum_{1}^{n} T_i.
\end{equation}

\subsubsection{Experimental Results} \label{sec:results2}
In this section, we introduce the experimental results of our uncertainty-aware tuning on the YCSB-a and TPCC workload.

\noindent\textbf{Latency Distribution \& Time Efficiency.} As shown in Figure~\ref{fig:tuningbar}, we present the 90-th percentile tail latency distribution and the average time per iteration of different tuners on the YCSB-a and TPCC workload. 

\begin{figure}[ht]
\centering
\subfigure{
    \includegraphics[width=\linewidth]{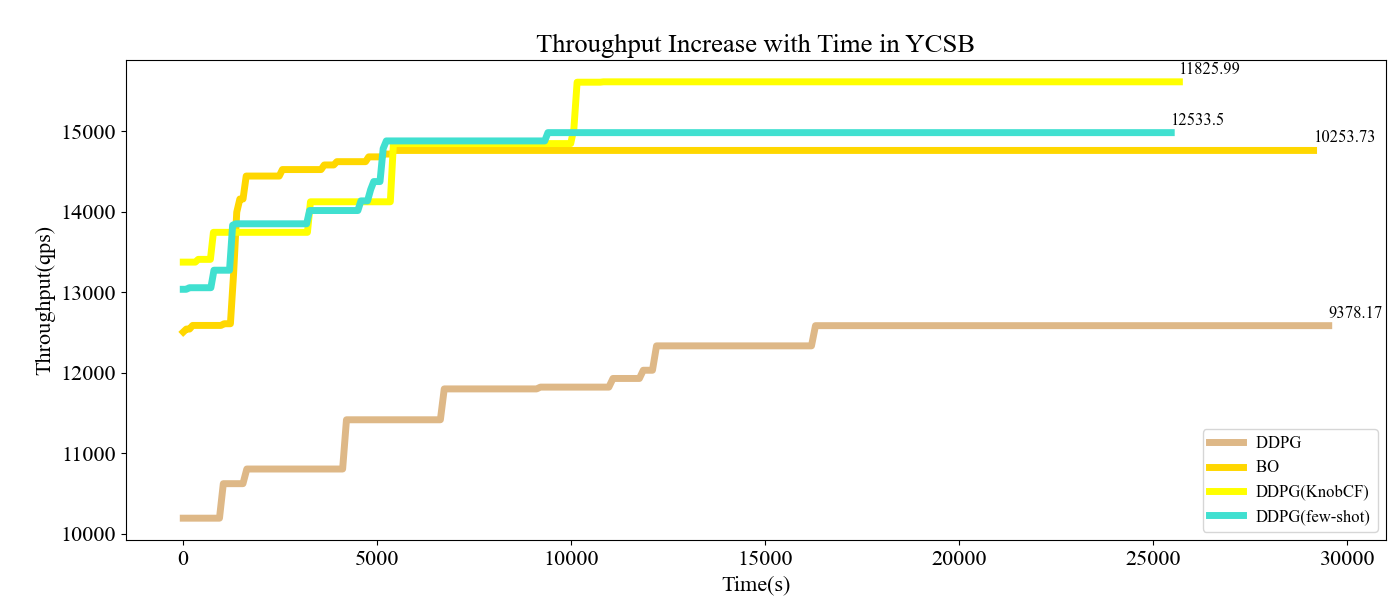}\label{fig:ycsb-line} 
	}
    \subfigure{
		\includegraphics[width=\linewidth]{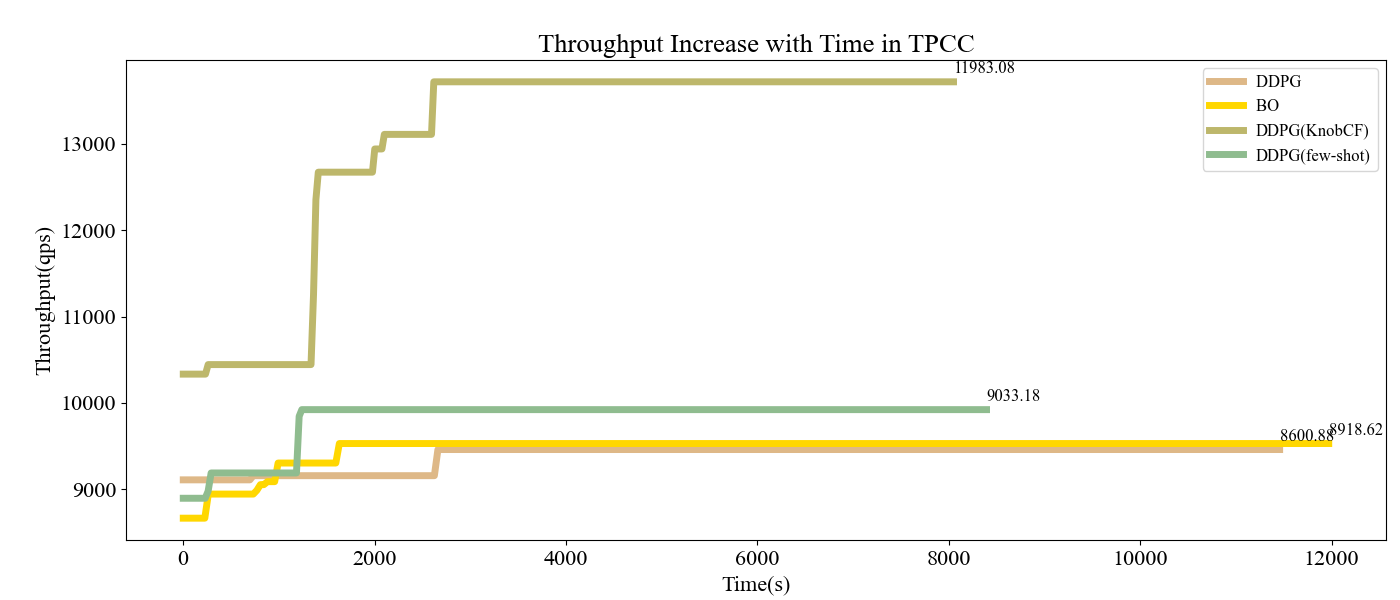}\label{fig:tpcc-line}
	}
    \caption{The maximum throughput trends over time}
    \label{fig:tuningline}
\end{figure}

On the YCSB-a, the KnobCF(few-shot) is pre-trained on historical tuning tasks of TPCH, TPCC, and Job-light, and fine-tuned by the evaluations of the former 30 iterations in YCSB-a. Then, we integrate the KnobCF and KnobCF(few-shot) to DDPG optimizer to implement the knob tuning in YCSB-a. For workload latency, we observe 82ms, 80ms average latency in DDPG(KnobCF) and DDPG(few-shot), which outperforms the 110ms, 95ms average latency in DDPG and BO. This indicates that our KnobCF achieves more accurate latency estimation, bringing lower workload latency. For the average time per iteration, we observe that the DDPG(KnobCF) and DDPG(few-shot) achieve the best average time per iteration of 75 and 80 seconds, which is significantly better than the average time per iteration of 125 and 100 seconds of the DDPG and BO. This indicates that our KnobCF could quickly complete the latency prediction, bringing high time efficiency for practical tuning tasks. 

On the TPCC, the KnobCF(few-shot) is pre-trained on the historical tuning tasks of TPCH, YCSB-a, and Job-light, and finetuned by the evaluations of the former 30 iterations in TPCC. We observe similar effects, with the DDPG(KnobCF) and DDPG(few-shot) achieving the best average latency of 17ms and 22ms. The DDPG and BO achieve an average latency of 27ms and 24ms, respectively. Also, the DDPG(KnobCF) and DDPG(few-shot) achieve the best average time per iteration of 26s and 27s, respectively. Overall, our KnobCF effectively guides the knob tuning and largely saves the evaluation time of the tuning tasks.

\noindent\textbf{Maximum Throughput Trends Over Time.} Figure~\ref{fig:tuningline} shows the maximum throughput trends over time on the YCSB-a and TPCC workload. In YCSB-a, we observe that DDPG(KnobCF) and DDPG(few-shot) achieve competitive maximum throughput to DDPG and BO methods in lower time consumption ( approximately 80\%-85\%). This indicates that our KnobCF could effectively reduce the time consumption of knob tuning evaluations while maintaining the knob tuning results. In TPCC, we observe similar improvements, i.e. DDPG(KnobCF) and DDPG(few-shot) achieve competitive maximum throughput to DDPG and BO methods with only 60\%-70\% time consumption. This time efficiency benefits from our KnobCF, which could efficiently predict the query latency distribution based on historical evaluations instead of executing the queries repeatedly. Traditional knob tuners~\cite{end2end-ce}, on the other hand, typically repeat the execution of all queries without considering uncertainty distribution.

\subsection{The robustness analysis}\label{sec:robustness}
In this section, we evaluate the robustness of our uncertainty-aware knob classifier and uncertainty-aware knob tuning method varying the dimension of classifier output from 8 to 16. In our classifier, the output dimension is a crucial hyper-parameter, which directly determines the number of categories of query uncertainty and affects the quality and efficiency of knob tuning. 
\begin{figure}[htb!]
    \centering
    \includegraphics[width=\linewidth]{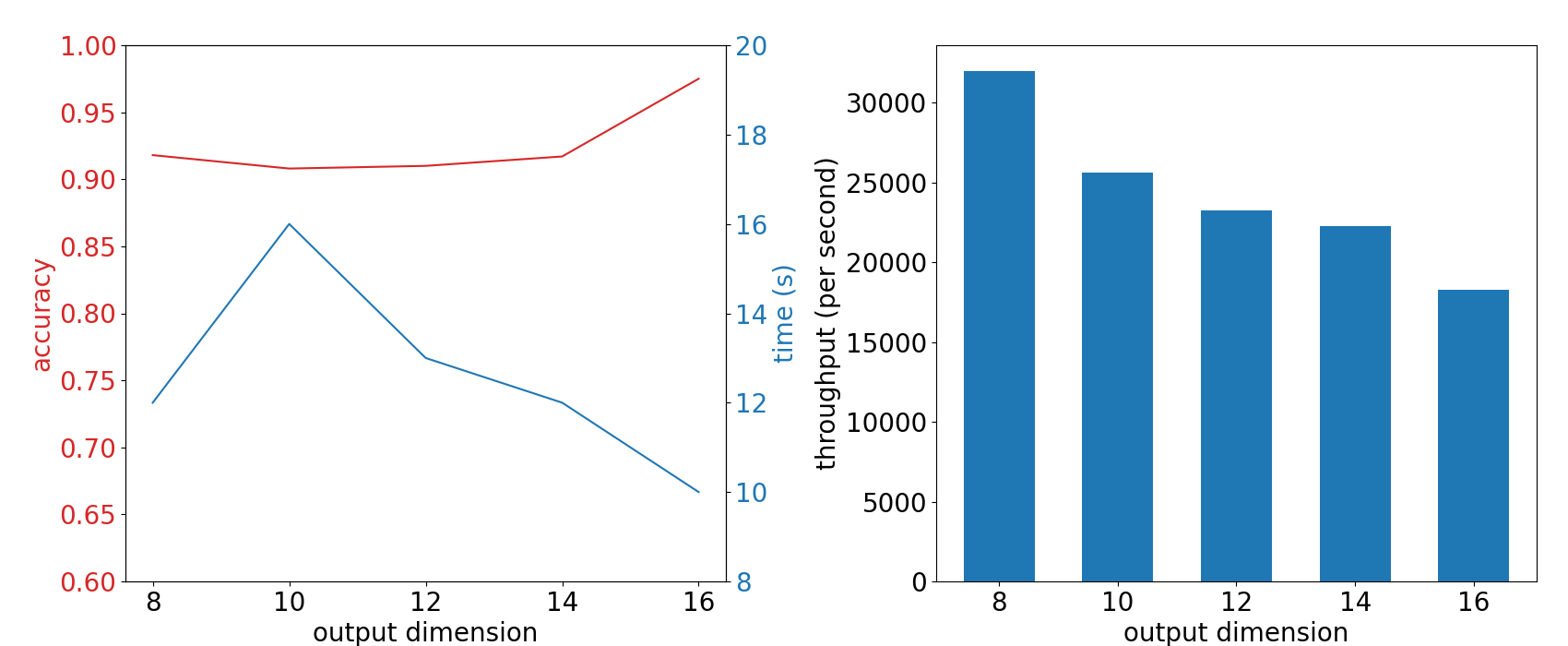}
    \caption{The effect of KnobCF(few-shot) in different output dimensions.}
    \label{fig:robusteff}
    \end{figure}

\subsubsection{The effectiveness of KnobCF under different output dimensions}
In this section, we conduct the robustness evaluation for KnobCF(few-shot) on YCSB-a. We utilize the TPCH, JOB-light, and TPCC to complete the pretraining of KnobCF and then utilize the former 30 iterations of YCSB-a tunings to finetune our knob classifier under different output dimensions of 8, 10, 12, 14, and 16. Similarly, we evaluate the model prediction accuracy, training time, and inference throughput of KnobCF(few-shot). 

\noindent \textbf{Prediction Accuracy:} As depicted in Figure~\ref{fig:robusteff}, the red line of the left figure presents the prediction accuracy changes with different dimensions in YCSB-a. We observe that our KnobCF(few-shot) maintains high prediction accuracy above 0.9 across various output dimensions. Even with the output dimension of 8, our KnobCF(few-shot) still achieves a prediction accuracy of 0.92, which is slightly lower than the output dimension of 16. This indicates that our KnobCF(few-shot) achieves stable and reliable prediction performance across various output dimensions. 

\noindent \textbf{Training Time:} As shown in Figure~\ref{fig:robusteff}, the blue line of the left figure presents the training time of KnobCF(few-shot) of different dimensions in YCSB-a. We observe that the average training time of KnobCF(few-shot) (13 s) is significantly shorter than the general KnobCF (160 seconds shown in Table~\ref{tab:query-plan-exp}). Specifically, in any dimension, the few-shot training time is under 20 seconds, accounting for less than 0.1\% of the total knob tuning time. This is because our KnobCF utilizes the former 30 evaluations of knob tuning to complete model training without any extra time consumption of collecting training data. The stable training time makes our KnobCF(few-shot) more efficient and practical in practical knob tuning tasks. 

\noindent \textbf{Inference Throughput:} The right figure in Figure~\ref{fig:robusteff} presents the uncertainty category inference throughput of KnobCF(few-shot) of different dimensions. Although the inference throughput decreases with the output dimension increasing, our KnobCF(few-shot) still achieves high inference throughput above 17000 queries per second. Also, this inference time is significantly lower than the query execution time, which could be further accelerated by GPU. 

\begin{figure}[htb!]
    \centering
    \includegraphics[width=\linewidth]{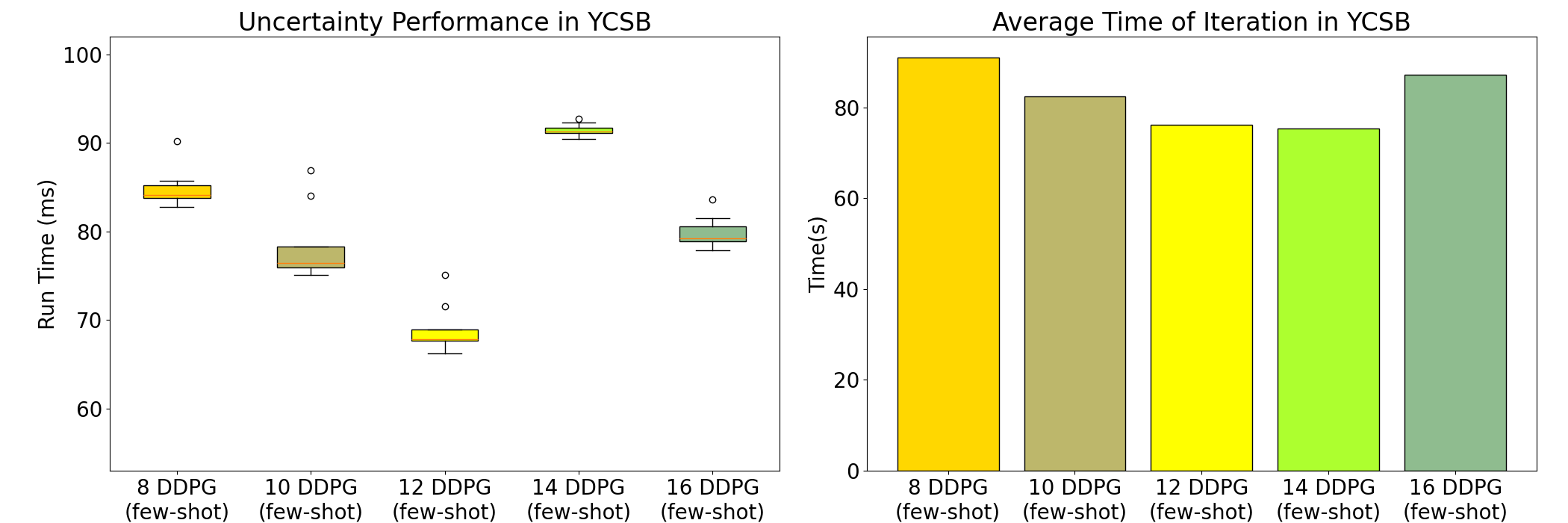}
    \caption{The Effect of DDPG(few-shot) in Different Output Dimensions.}
    \label{fig:output-bar}
    \end{figure}

\subsubsection{The Effect of Different Output Dimensions in Knob Tuning.}
In this section, we evaluate the tuning robustness of DDPG (few-shot) in YCSB-a under different output dimensions, including the uncertainty latency, average runtime of tuning iterations, and the maximum throughput trends over time. 

Figure~\ref{fig:output-bar} shows the uncertainty workload latency of optimal knob configuration and average runtime of tuning iterations of the DDPG (few-shot) with output dimensions of 8, 10, 12, 14, and 16. We could observe that DDPG (few-shot) achieves similar uncertainty workload latency, 78ms in dimension 10 and 79ms in dimension 16. Also, the average runtime of tuning iterations is stable across different output dimensions. This indicates that our KnobCF(few-shot) makes stable uncertainty predictions to enhance the knob tuning.

\begin{figure}[htb!]
    \centering
    \includegraphics[width=\linewidth]{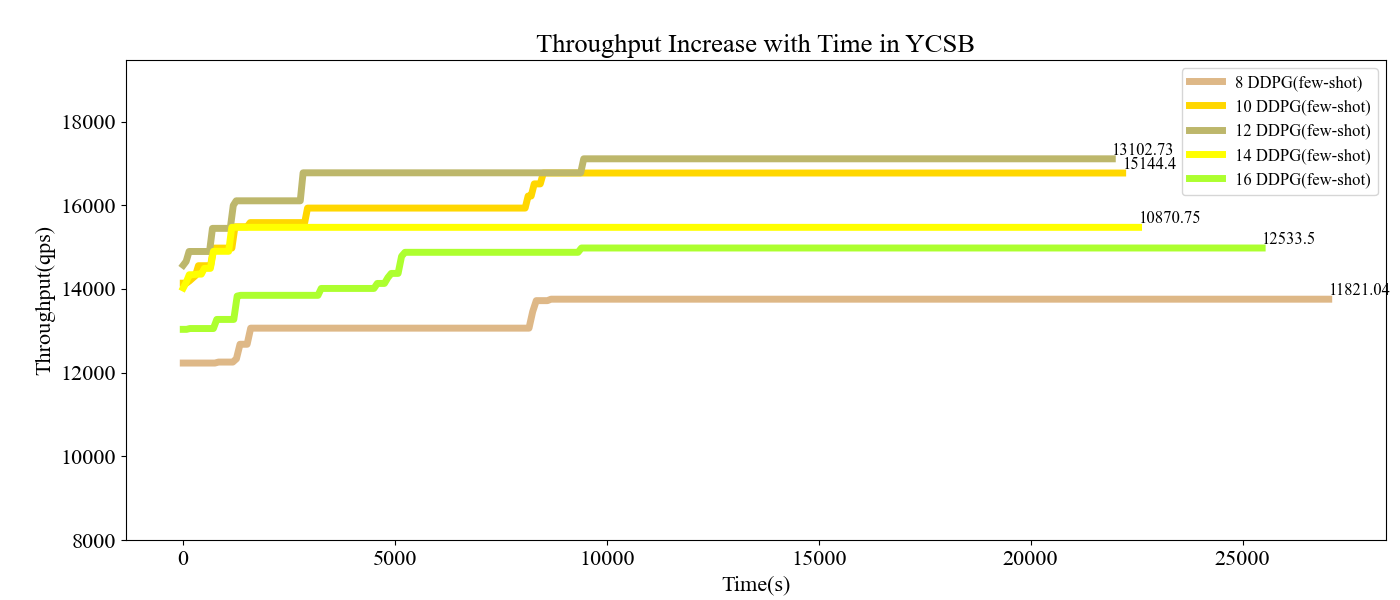}
    \caption{Maximum throughput trends over time in different output dimensions.}
    \label{fig:output-line}
    \end{figure}

Figure~\ref{fig:output-line} shows the throughput increase of the DDPG (few-shot) with the time consumption under different output dimensions. We observe similar tuning trends in different output dimensions. This indicates that limited output dimensions are sufficient to represent the uncertainty distribution of knob configurations and make effective knob tuning.

\section{Related work}
In this section, we introduce existing works from two aspects, including the knob tuning methods and the query cost estimation methods.
\subsection{Knob Tuning Mehods}
With the development of cloud databases, automatic knob tuning plays a crucial role in database performance optimization. Existing automatic database knob tuning methods~\cite{zhao2023automatic} can be roughly classified into three categories: heuristic methods, Bayesian optimization methods, and reinforcement learning methods. Specifically, Sullivan et al.~\cite{sullivan2004using} propose heuristic knob tuning, which involves random sampling of a small number of knob settings to automate the exploration of appropriate knob settings. Subsequently, researchers propose some Bayesian-optimizer-based knob tuners~\cite{van2021inquiry, tan2019ibtune}, such as Restune~\cite{zhang2021restune} and OnlineTune \cite{zhang2022towards}, which recommend suitable knob configurations through iterative sampling and evaluation of knob settings. 

Compared to BO-based tuning methods, reinforcement learning-based knob tuning methods do not require training data preparation. For example, Zhang et al. propose CDBTune \cite{zhang2019end}, which builds a reinforcement learning-based tuning model DDPG. During iterations, this model utilizes an agent to recommend tuning operations based on state features and updates the tuning strategy through rewards to optimize database performance. Li et al. \cite{li2019qtune} proposed Query-based reinforcement learning tuning, Qtune, integrates query and workload features into the reinforcement learning tuning model, allowing better adaptation to new scenarios. 

Further, to reduce tuning time, some improved reinforcement learning methods have been proposed. For instance,  Wang et al.~\cite{wang2021udo} propose UDO, a universal database optimization based on Reinforcement Learning, which reduces the number of database restarts during tuning knob configuration by reorganizing the order of evaluations. Cai et al. \cite{cai2022hunter} propose HUNTER to preemptively reduces the configuration search space through knob selection and genetic algorithm, to further largely reduce the time consumption of knob tuning. 

Additionally, DB-BERT proposed by Trummer et al. \cite{trummer2022db} uses a pre-trained BERT language model to extract useful tuning methods from database manuals and search engines, combined with a simpler DDQN algorithm to select knobs and recommend tuning.

In summary, existing knob tuning methods have made great progress in reducing the knob candidate space and making effective recommendations. However, these methods still face the challenges of useless evaluations, overestimation, and underestimation. In this paper, we consider to address these challenges from the aspect of the query uncertainty distribution.

\subsection{Query Cost Estimation}
Query cost estimation~\cite{end2end-ce} occupies an important position in database management systems, as it is a key part of database tuning strategies, playing a vital role in query optimization, index optimization, storage efficiency, etc. Early on, to improve the efficiency of query optimization, researchers propose some statistical methods to estimate query costs hypothetically. For example, Li et al.~\cite{li2012robust} propose an operator-based statistical technique to estimate query execution time. 

Recently, the database community has attempted to use deep learning models \cite{sheoran2022conditional,yu2022cost} to improve the accuracy of query cost estimation. For example, Sun et al. propose an end-to-end query cost estimation framework \cite{end2end-ce}, which uses a tree structure model to learn the relationships between queries and performance labels. Marcus et al.~\cite{Tree-Rnn} analyze the query plan tree structure and propose a plan structural deep neural network for query cost estimation. Hilprecht et al. designed a zero-shot cost model~\cite{hilprecht2022zero} for query cost estimation, which has the advantage of being able to generalize to unseen databases. 

Furthermore, some studies have proposed uncertainty cost estimation methods. For example, Wu et al.~\cite{wu2014uncertainty} introduce the concept of query uncertainty, suggesting that the performance of queries can be represented by a distribution due to runtime influences (e.g., random access). Similarly, Dorn et al.~\cite{dorn2020mastering} also propose that configurable software systems exhibit performance estimation uncertainty. Different from existing query cost estimation methods, we consider the query uncertainty estimation from the perspective of knob configurations.

\section{Conclusion}
\label{sec:conclusion}
KnobCF effectively tackles the useless evaluations and the overestimation and underestimation issues in the traditional knob tuning framework. It is a general framework that can be applied to various DBMSs and tuning tasks. For useless evaluations, we propose the uncertainty-aware knob classifier to predict the category label of a certain knob configuration according to historical evaluations. For the overestimation and underestimation, we model the joint performance distribution of knob configurations instead of the single-point estimation. Importantly, KnobCF is based on transferable feature representation learning, bringing effective model transfer for different tuning tasks. The experimental results demonstrate that KnobCF significantly reduces the knob tuning time and improves the tuning efficiency compared to the state-of-the-art methods. 

\begin{acks}
This work was supported by the National Natural Science Foundation of China (No.~62232005, No.~62202126, No.~92267203).
\end{acks}

\clearpage

\bibliographystyle{ACM-Reference-Format}

\bibliography{main}

\end{document}